# Highly sensitive ammonia sensor using reflection of light at a glass - photonic crystal interface

A. S. Kuchyanov [1,]*, P. A. Chubakov [1,], H. Spisser [2,] and A. I. Plekhanov [1]

[1] Institute of Automation and Electrometry SB RAS, 1 Prosp. Acad. Koptyug, Novosibirsk 630090, Russia;

[2] Institut d'Optique Graduate School, Campus Polytechnique Address line 1, 2 Avenue Augustin Fresnel, Palaiseau cedex 91127, France;

* *Author to whom correspondence should be addressed; E-Mail: aleks@iae.nsk.su ;*
*Tel.: (383)3306655.*

We have discovered and studied the effect of the asymmetric deformation of a photonic crystal in the form of a change in the slope of the crystal planes as it is filled with a gaseous analyte. We have demonstrated that the use of a new effect leading to the displacement of the stop band against the unchanged spectrum of diffracted white light at the (glass–thin opal film) interface can be used as fast, compact, high sensitive and reproducible optical chemical sensor for ammonia. Low cost and simplicity of sensor fabrication, the response of which can be easily observed without resorting to spectral instruments are therefore likely to be attractive. The basis for high sensitivity (1 ppm), fast response (120 ms) is capillary vapor condensation. On the basis of this effect a cheap high-speed and highly sensitive gas sensors has been built.

## 1. Introduction

Optical techniques for the detection of liquid and gaseous analytes are interesting in a wide range of applications. While optical spectroscopic sensors are mature and widely used, emerging alternative optical technologies are significantly contributing to advancements in the current sensing techniques.

There is a high interest for portable, low cost, fast and reliable gas sensors in a number of industrial and biomedical areas such as point-of-care health monitoring and light-weight industrial sensing [1]. However, due to various limitations in current detection technologies, gas sensors satisfying these requirements are barely available today. Ammonia ($NH_3$) detection draws a considerable interest with regard to environmental, industrial, explosives detection and medical perspectives. The allowed exposure limit in such environments is about 20 ppm. On the other hand, breath ammonia has been recognized as a bio-marker for human physiological disorders. For example, a high correlation between blood urea nitrogen and breath ammonia level in kidney patients has been reported [2]. Interesting material properties, such as a high internal surface area, tunable refractive index and biocompatibility, make photonic crystal (PhC) a strong candidate for a material to be used in various applications [3-5].

Their small size, high sensitivity, and resilience in hostile environments give them a distinct advantage over traditional electronic sensors. There are numerous sensor applications of PhCs based on the measuring of its spectral characteristics (the peak positions of photonic band gaps (stop bands) depend on the filling of pores with different substances. This dependence is caused either by a change in the effective refractive index of PhC or a changes of the lattice constant during vapor condensation [6]. In most cases the monitoring of the change of optical properties, such as wavelength shift of the optical stop bands or the Fabry-Perot fringes of the PhC occurs when the light is incident on the air - PhC interface.

We report the results of the investigation of a new effect of the displacement of the stop band against the unchanged spectrum of diffracted white light at the (glass–thin opal film) interface and the possibilities of using this effect to create fast and high sensitive optical chemical sensors.

## 2. Sensing with photonic crystal

In the beginning we consider the rays of light incident on the interface between glass and PhC, illustrated in Fig. 1.

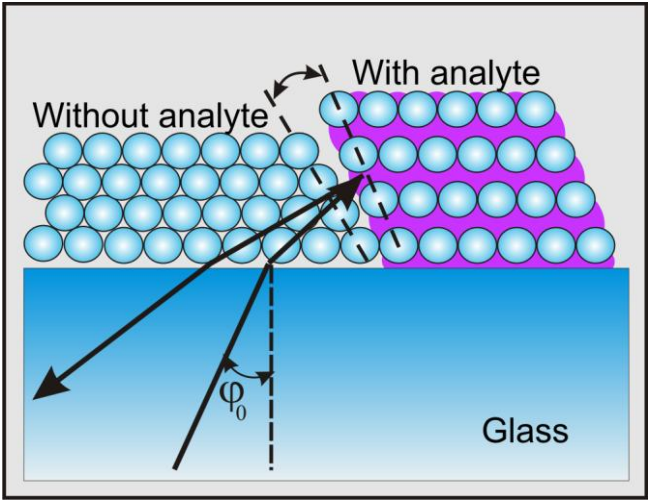

FIG.1. Illustration of the change in the slope of the PhC crystal planes due to filling with the analyte.

The PhC is a film consisting of many layers of monodisperse spherical silica particles (MSSP) packed in a face-centered cubic crystal structure.

The angular spectrum of light reflected from the interface between glass and PhC is made up of several components: the diffraction spectrum by reflection from the boundary layer of the MSSPs, and the spectrum of the Bragg reflection from the crystal planes $(\bar{1}11)$, $(1\bar{1}1)$, $(11\bar{1})$ at an angle of $60^0$ to the substrate plane (band gap). The spectrum of the transmitted light is described in our paper [7].

The rays diffracted by the glass-PhC interface propagate back to the glass according to the formula:

$$\lambda = d \cdot n_g (\sin\varphi_0 + \sin\varphi_1) \quad (1)$$

where $d$ is the diameter of the MSSP, $n_g$ is the refractive index of the glass prism, $\varphi_0$ is the angle of incidence on the glass - PhC interface, $\varphi_1$ is the diffraction angle of the reflection.

The rays which refract at the glass-PhC interface are reflected from the crystal planes in accordance with the Bragg condition:

$$\lambda = 1.63 d \sqrt{n_{eff}^2 - \left(\sin^2\left(60 - \arcsin\left(\frac{1{,}52}{n_{eff}}\sin\varphi_0\right)\right)\right)} \quad (2)$$

$$n_{eff} = \sqrt{n_1^2 \cdot \beta + n_2^2 (1-\beta)} \quad (3)$$

where $n_1$ is the refractive index of silica, $n_2$ is the refractive index of the substance filling the pores of the PhC; $\beta = 0.74$ is the PhC fill factor.

As a result, the reflection of white light from the glass-PhC interface leads to an intense peak on a background of uniform spectral-angular distribution as shown in Fig. 2a.

We have seen a significant shift (up to $5^0$ or 10 nm across the spectrum) of the maximum intensity in the angular distribution of the reflected light when vapor of polar molecules filled the PhC. It is essential that the shift can easily be observed without resorting to spectral instruments. In this manuscript, we report the nature of this effect, on which a highly sensitive optical-chemical sensor can be fabricated.

In the first place, according to the Equation 2 pour filling of the PhC by a gas, for example, such as $NH_3$ ($n_2 = 1{,}000375$) instead of air ($n_2 = 1{,}00028$) changes the wavelength of the stop band of light reflected from the PhC by only 0.0074 nm. Most likely, filling the pores of the PhC with certain molecules leads to the change of the lattice constant of the PhC. Let us consider the reasons that could lead to such change. It is well known that siloxane (Si–O–Si) and silanol (Si–OH) groups are present on the surface of silica nanoparticles [8]. The presence of a mobile hydrogen atom in polar hydroxyl groups gives rise to the effective interaction with the molecules of the gas and liquid phases by two ways. The first interaction is electrostatic attraction between hydroxyl groups on the silica surface and the surrounding dipole molecules. In addition, hydrogen bonds can appear between hydroxyl groups and the surrounding molecules if these molecules have an undivided electron pair. For this reason, water and ammonia, being the most polar molecules among those listed above, manifest the strongest electrostatic interaction and form the strongest hydrogen bonds with the silica surface with a high probability of the formation of monolayers or cluster island films on the surface of nanoparticles. Indeed, the experiments show that the maximum spectral shift, about 10 nm, of the stop band of the PhC is observed under the action of water or ammonia vapors. A smaller spectral shift of 1-2 nm was observed for ethanol and isopropyl alcohol vapors and was absent for carbon tetrachloride vapors; this observation corresponds to a decreasing sequence of the dipole moments of these molecules [9]. It should be noted that the revealed selectivity can be enhanced by the controlled chemical modification of the surface of silica nanoparticles as a carrier [10]. Adsorption of analyte molecules on the surface of MSSPs leads to capillary condensation, as a result of which may be a change of the distance between MSSPs [11]. We found experimentally that such a change occurs only in the direction across the PhC film. Fig. 1 illustrates how a change in the distance only between the MSSPs in the transverse direction can change the slope of the crystal planes $(\bar{1}11), (1\bar{1}1), (11\bar{1})$, and, as a consequence, induce a shift of the peak of the angular distribution of the output radiation.

The effect of swelling PhC consisting of polymer particles is well known [12-14]. The PhC material consists of a hydrogel, which embeds a crystalline colloidal array of monodisperse polystyrene spheres. In this case swelling of the polymer film during exposure to specific gases, lead to a change in total effective refractive index or/and to a change in the lattice constant of the PhC. Most likely, in this an anisotropic strain of the PhC should not be observed. However, this thesis requires further experimental check. In our case, the analyte is mainly on the surface of MSSPs.

The effect described above is encouraged to use for the development of relatively cheap sensitive optical sensor for various liquids and gases without additional spectral instruments. In this case, PhC serves simultaneously as a spectral instrument and an optical chemical sensor. The reversible structural change of self-assembled three-dimensional colloidal PhCs facilitates the development of reproducible and reusable vapor sensors that can be operated over wide range of temperatures.

## 2. Experimental Section

### 2.1. Colloidal Dispersions

Silica spheres of 260 nm diameter were synthesized using a Stöber process [16]. Coagulation-resistant nanosize MSSP suspensions are obtained by the hydrolysis of tetraethoxysilane ($Si[OC_2H_5]_4$) in an alcohol solution in the presence of ammonia playing the role of the process catalyst and, then, of an electrolyte, which is a potential-forming and suspension-stabilizing factor.

### 2.2 PhC Film Preparation

Films of PhC were deposited onto glass prism from ethanolic colloidal dispersions using the vertical assembly method [17]. The glass prism washed with distilled water and then ultrasonicated for 30 min in carbon tetrachloride. The prism faces were finally rinsed with distilled water and dried with a dry air stream. Then the prism was placed in MSSP suspension. The prism faces were located vertically. The dispersions were left in a furnace at 45 °C overnight so all the ethanol evaporated. Nanocrystalline opal films has a brilliant "varnish" appearance, and cause a bright homogeneous diffraction of incident light. Their thickness can be controlled from 0.5 to 2–8 nm (from 2–3 to 30 MSSP layers) by varying the MSSP concentration in the suspension.

### 2.2 Characterization

Such a film represents a single crystal over the entire area, which is corroborated by scanning electron microscopy examination. The MSSPs in the film are stacked in hexagonal close-packed layers corresponding to the (111) plane of the face-centered cubic lattice and are parallel to the substrate surface. The film surface area reaches 1–2 $cm^2$. Spectral measurements were taken by a spectrometer (AvaSpec-2048 TEC)

## 3. Results and Discussion

To study the spectral - angular characteristics of the white light, which diffracted on the glass - PhC interface, a film of opallike PhC has been grown on one of the faces of a glass prism. The film thickness was determined 30 to be layers of MSSPs with a diameter of 260 nm. The PhC film has a minimum number of defects in a large area of 15x25 mm. The boundary of the film growth is chosen parallel to the plane of incidence. To provide a wide range of angles of incidence on the glass - PhC interface a glass prism was used because the effect was observed for angles of incidence greater than $40^0$. The prism coated with a PhC film was placed in a sealed chamber having a transparent window for input and output of radiation. Next $NH_3$ gas pre-mixed with air at the appropriate concentration was released into the chamber. We have also used water and ammonia vapors. We exposed the PhC device to different parts-per-million (ppm) of $NH_3$ diluted in air at atmospheric pressure and room temperature (~300 K) for a few seconds. The concentration of ammonia was measured by a commercial sensor TGS 2444 and water vapor was measured by a moisture meter. A collimated beam of white light from a halogen lamp (2) incident on the PhC has a diameter of 2 mm. The reflected diffracted light is directed through a fiber to the spectrograph «AvaSpec-2048 TEC», with a resolution of 0.7 nm. The fiber was placed on a goniometer arm.

We carried out the experimental verification of the anisotropic change in the distance between the MSSPs of the PhC during vapor condensation. We have used a narrowband laser beam as a light source with a wavelength that does not coincide with the PhC band gap to prove the invariance of the distance along the film. In this case, the angular spectrum of the reflected light contains only the component that depends on the diffraction on the first layer of MSSPs along the PhC film. It was found that the angular position of the laser radiation is not changed. It denotes the constant distance between the MSSPs along the PhC film when the sensor is exposed to the analyte.

All reflectance spectra obtained from PhCs film display well-defined Fabry-Perot fringes, indicating high cystalline quality of the fabricated colloidal PhCs. The change of the distance between the MSSPs in the direction across the PhC film was checked using Fabry-Perot fringes

We have estimated that the film thickness change was 470 nm, and the change in the distance between MSSPs across the film is 17 nm.

In addition, further evidence of the formation of the optical anisotropy during the condensation of analyte vapor in the PhC film is the observation of the induced birefringence [15]. In this experiment white light is passed through rossed polarizers between which the PhC film is placed. The offset of isogyre in the conoscopic patterns clearly indicate that the optical anisotropy in this case is caused by the increase of PhC film thickness in the transverse direction.

The illustration of the sensor operation for two different concentrations of ammonia vapor is shown in Fig. 2b.

The light diffracted at the glass-PhC interface arrives on two photodetectors (PDs) placed at the angles $\varphi_1$ and $\varphi_2$ corresponding to the rise and fall of the curve plotted in Fig. 2a. The signals from them change in opposition because of the curve shift during anlalyte vapor condensation. The PD's signals arrive at the amplifier of signal ratio. This measurement circuit eliminates the fluctuations in the intensity of the incident light and increases the signal response of the system.

In our experiments, humidity was measured separately and its effect was treated as a constant parameter in our measurements. So it allows us not to drain the camera before measurement.

Fig. 3a shows that the PD signal is nearly linear with the logarithm of ammonia partial pressure. With a halogen lamp as a light source a minimum detectable level of ammonia is found about a few ppm and the maximum detected amount of more than 1000 ppm is obtained for the saturated vapor at room temperature.

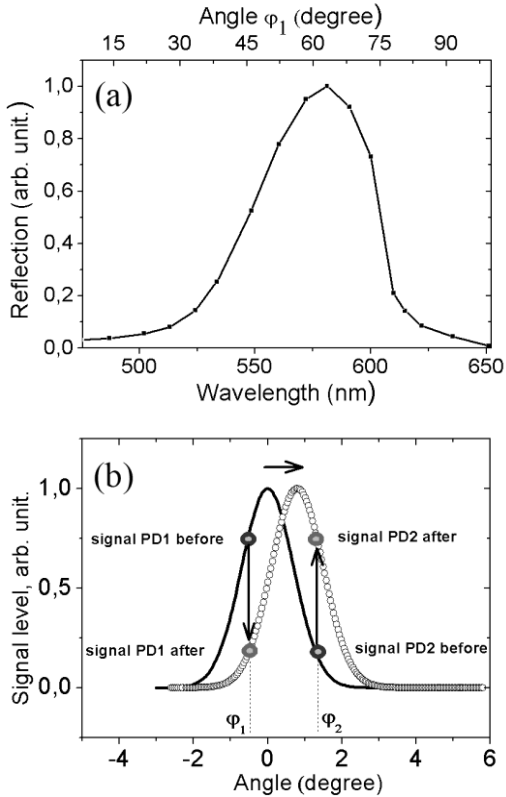

FIG.2. a) Reflectance spectrum obtained from the interface of the glass – PhC film.
b) Illustration of the operation of the sensor for two different concentrations of ammonia vapor. The vertical arrows show the change in the signal level of the PD1 and PD2 when exposed to the analyte. $\varphi_1$ and $\varphi_2$ are the PD angular positions.

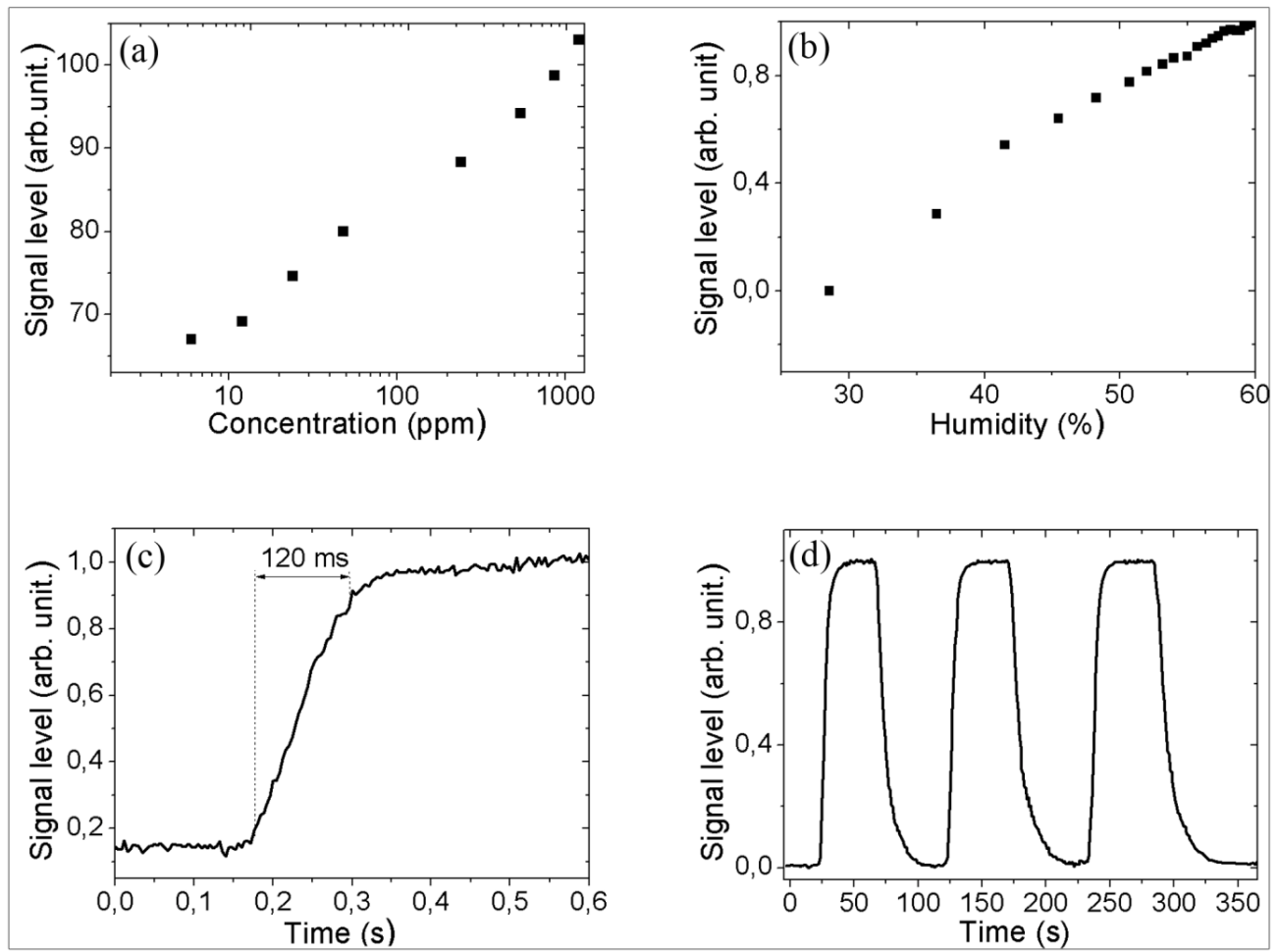

FIG.3. Response of the PhC sensor. a) Dependence of the PDs signal on the concentration of ammonia vapor. b) Dependence of the PDs signal on the concentration of ammonia vapor. c) Temporal response of the PhC sensor. d) Reversibility of the PhC sensor for detection of ammonia. Normalized change in electric signal of the PDs as a function time for detection of 1000 ppm of $NH_3$ in three cycles. The adsorption and desorption steps are performed at room temperature.

Importantly, in order to significantly improve the detection sensitivity we used a laser as a light source which wavelength may be different from the stop band. With a laser with a narrow spectrum this approach leads to an angular width of the curve of $1.5^0$ what is a sharp steepness for the slopes, and the sensitivity of the sensor can be improved by nearly two orders of magnitude.

Fig. 3b shows the contribution of moisture to the measured signal of PD. The signal is proportional increased with the relative humidity.

Fig. 3c shows the temporal response of the sensor to a rapid increase of the concentration of ammonia. In this case a stream of air with ammonia vapors at a high pressure was directed toward the PhC film. To eliminate the effect of pressure on the signal we carried out a check of the influence of a jet of dry air with the same pressure without ammonia. Thus, the response time of our sensor was 120 ms, with a recovery time of about 10 s. In our case,
the response time of the sensor is largely limited by the filling time of the chamber with various gases.

Our sensors enable the reversible and reproducible detection of large range of ammonia pressure. Fig. 3d shows the reversibility of the PhC sensor over several cycles during adsorption and desorption steps. This figure shows the normalized electric signal of PDs as function of time for different concentration of $NH_3$ ranging from 1000 ppm to 0 ppm. All tests are performed at atmospheric pressure and room temperature. Typical $NH_3$ PhC gas sensor operates in the 10 to 1000 ppm range of $NH_3$ concentration in air.

4. Conclusions

In conclusion, we have demonstrated that the use of a new effect leading to the displacement of the stop band against the unchanged spectrum of diffracted white light at the (glass–thin opal film) interface can be used as fast, compact, high sensitive and reproducible optical chemical sensor for ammonia. Low cost and simplicity of sensor fabrication, the response of which can be easily observed without resorting to spectral instruments are therefore likely to be attractive. The detection sensitivity can be greatly improved by using a laser with a narrow spectrum as a light source. The basis for high sensitivity (1 ppm), fast response (120 ms) is capillary vapor condensation, which increases the distance between MSSPs in the direction across the PhC film. Such asymmetric deformation causes a change in the slope of the crystal planes from which the incident beam is reflected. The angle changing of the reflected beam is recorded as a signal which is proportional to the concentration of ammonia. In addition to ammonia, the vapor of other polar molecules can act as the analyte. The functionalization of MSSP surface allows to fabricate other gas sensors using the principle which we have described.